\documentclass[preprint,showpacs,preprintnumbers,amsmath,amssymb]{revtex4}
 \usepackage{graphicx}
 \usepackage{dcolumn}
 \usepackage{bm}
 \newcommand{\R}{\mathbb R}
 \newcommand{\C}{\mathbb C}
 \newcommand{\HH}{\mathfrak{H}}
 \newcommand{\HD}{\hat{\mathcal{H}}}
 \newcommand{\N}{\mathcal{N}}
\begin{document}
 \title{The Construction of the Dual Family of Gazeau-Klauder Coherent
   States via Temporally Stable Nonlinear Coherent States}
 \author{R. Roknizadeh}
 \email{rokni@sci.ui.ac.ir}
 \author{M. K. Tavassoly}
  \email{mk.tavassoly@sci.ui.ac.ir}
\affiliation{Quantum Optics Group, Physics Department, University
             of Isfahan, Isfahan, Iran}
\begin{abstract}
  Using the {\it analytic representation} of
  the so-called Gazeau-Klauder coherent states(CSs),
  we shall demonstrate that how a new class of
  generalized CSs namely the {\it family of dual states} associated with
  theses states can be constructed through viewing these states as
  {\it temporally stable nonlinear CSs}. Also we find
  that the ladder operators, as well as the displacement type operator
  corresponding to these two pairs of generalized CSs,  may be easily obtained
  using our formalism, without employing the {\it supersymmetric
  quantum mechanics}(SUSYQM) techniques.
  Then, we have applied this method to some physical
  systems with known spectrum, such as P\"{o}schl-Teller, infinite
  well, Morse potential and Hydrogen-like spectrum as some quantum mechanical systems.
  Finally, we propose the generalized form of Gazeau-Klauder CS and
  the corresponding dual family.
  \end{abstract}
\maketitle
\section{{\bf Introduction}}\label{sec-intro}
  Coherent states(CSs) play an important role in various fields of
  physics, quantum technologies and especially in
  quantum optics(see for instance \cite{Klauder1985,Chong,Ali2000,Deuar}).
  Therefore efforts along generalizations and applications have been
  appreciably increased  in recent years \cite{Rokni2004}.
  Recently, Gazeau and
  Klauder have introduced an important class of generalized CSs,
  the so-called  {\it "Gazeau-Klauder CSs"} have been denoted by
  $|J,\gamma \rangle$, corresponding to any arbitrary quantum
  mechanical system \cite{Gazeau1999,Gazeau-Monceau2000}.
  Keeping in mind that Gazeau-Klauder CSs are really CSs,
  we will refer to them as {\it "GK states"}.
  These states have attracted much
  attention in literature (see e.g.
  \cite{Roknizadeh2004,Popov2003,Antoine2001,El Kinani2001,
  El Kinani2002,El kinani2003}).
  More recently, along generalization of GK states, the vector CSs of
  the GK type have been constructed and some physical applications
  of them have been addressed \cite{ali-bag}.

  In another direction, a new way of
  generalization has been proposed to construct the so-called {\it family of the
  dual states} corresponding to some particular known classes of
  CSs such as nonlinear CSs \cite{Roy-Roy2000,Ali2004}. In an extended framework
  we have recently studied this idea,
  re-derived basically and well developed \cite{Ali2004}.
  The construction of the dual pairs may be actually performed for all
  classes of generalized CSs, obtained by each of the three ways of
  generalizations; i.e., {\it algebraic, symmetric and dynamics}.
  We have found that the canonical CS are
  the only {\it self-dual} family (a useful check point for our
  future construction in this paper). Also we
  have already established the dual states associated with
  KPS \cite{Klauderetal2001},
  PS \cite{Penson1999} and $SU(1, 1)$ group
  CSs successfully, of course after demonstrating the
  nonlinearity nature of these states \cite{Roknizadeh2004}.
  Unfortunately, as we have stated in earlier works, employing
  either our previous approaches in \cite{Ali2004} or the
  formalism proposed in \cite{Roy-Roy2000} for constructing the dual
  of GK state, do not lead to full consistent CSs with Gazeau-Klauder's criteria.
  For instance according to the proposition in \cite{Ali2004}
  the $\hat T^{-1}$-operator, whose action on
  canonical CS yields $|J, \gamma\rangle$ may be obtained trivially. But when one acts
  the related $\hat T$-operator on the standard CS, the output state which expected to
  be of the GK type (in fact the dual set of GK states), encounters some difficulties. For
  example, apart from the ill-definition of the $\hat T$ (and so $\hat T^{-1}$)-operator,
  the obtained states do not fulfill {\it temporal stability}.
  Consequently, the important
  property of the GK states, {\it the action identity} evolves some problems.
  Therefore one has to try some {\it radically different method} from the previous ones.

  This article organizes as follows:
  firstly, after imposing a {\it second modification} on the
  {\it modified} GK states $|z, \alpha\rangle$({\it to
  be distinguished from "GK states", by abuse of notation we shall call them as GKCSs})
  introduced by El Kinani and Daoud \cite{El Kinani2001},
  we shall clarify the nonlinearity of these
  states. Secondly, in view of establishing the GKCS as an extension
  of {\it "KPS nonlinear CS"} \cite{Klauderetal2001} to {\it "temporally stable CS"},
  together with the fact that the dual family
  of KPS nonlinear CS has already been introduced appropriately,
  we attempt to find the dual of GKCS,
  (we shall refer to it as DGKCS)
  through generalization of "the dual of KPS nonlinear CS" to
  state that it possesses the {\it "temporal stability"} characteristic.
  Upon generalizing this result, we shall introduce the
  $\hat{S}(\alpha)$-operator $(\alpha \in \mathbb{R})$
  which transfers any generalized nonlinear CS,
  corresponding to a Hamiltonian with known spectrum (which does not preserve
  temporal stability) to a situation in which it nicely restores this property.

  Additionally, a set of new interesting results such
  as the explicit form of annihilation, creation and displacement type operator
  corresponding to each of the two generalized CSs (GKCSs and DGKCSs) will be obtained.
  Also, by using the dual family of GKCS, the even, odd and Schr\"{o}dinger
  cat CSs  have been introduced.
  We then apply the method to some well-known solvable systems,
  i.e., harmonic oscillator, P\"{o}schl-Teller, infinite well and Morse
  potential and Hydrogen-like spectrum as some examples of quantum mechanical systems.
  Finally, we outline a scheme for {\it generalization of the GKCSs} as well as {\it DGKCSs}.
\section{\bf Analytical representations of Gazeau-Klauder
  CS as nonlinear CS}\label{sec-GKCS-NL}
  In this section,
  we first revisit the analytical representations
  of GKCSs and then impose a second modification on them. At last, we
  establish their nonlinearity nature.
\subsection{\bf Analytical representations of GK states}
   \label{sec-GKCS-NL}
  The GK states, $|J, \gamma\rangle$,
  corresponding to any Hamiltonian with discrete (non-degenerate) eigenvalues $e_n \geq
  0$, are defined as  \cite{Gazeau1999,Gazeau-Monceau2000, Klauder1998}:
 \begin{equation}\label{GKCS}
   |J, \gamma \rangle \doteq \N(J)^{-1/2}
   \sum_{n=0}^{\infty}\frac{J^{n/2}e^{-i \gamma e_n}}{\sqrt{
   \rho(n)}}|n\rangle,  \qquad  J\geq 0, \qquad  -\infty < \gamma <
   \infty,
 \end{equation}
   where $\N$ is a normalization constant may be determined
   (for a new and interesting formalism related to degenerate Hamiltonian
   see Ref. \cite{ali-bag}).
   The orthonormal set  $\{|n\rangle\}_{n=0}^\infty$ satisfy the eigenvalue equation:
 \begin{equation}\label{kps-14}
   \hat{H}|n\rangle=E_n|n \rangle \equiv \hbar \omega e_n |n
   \rangle = e_n|n\rangle, \qquad \hbar \equiv 1,
   \qquad \omega\equiv 1.
 \end{equation}
    The eigenvalues of the Hamiltonian $\hat{H}$ are such that
 \begin{equation}\label{Ineq}
    0 = e_0 < e_1 < e_2 < \cdots <e_n < e_{n+1} < \cdots.
 \end{equation}
   These states should satisfy the
   following properties: i){\it continuity of labeling}, ii){\it resolution of the
   identity}, iii){\it temporal stability} and iv){\it action
   identity}. The last two conditions requires $\rho(n)=[e_n]!$.

    Along the works on GK states, El Kinani and Daoud in a series of papers
    \cite{El Kinani2001, El Kinani2002, El kinani2003},
    imposed a minor modification on these states
    via generalizing the Bargman representation for the standard
    harmonic oscillator \cite{Bargman1961}.
    The authors
    introduced the {\it analytical representations of GK states},
    denoted by us as GKCSs:
\begin{equation}\label{GKED}
    |z, \alpha \rangle \doteq \N(|z|^2)^{-1/2}
    \sum_{n=0}^{\infty}\frac{z^{n}e^{-i \alpha e_n}}{\sqrt{
    \rho(n)}}|n\rangle,  \qquad  z \in \C, \qquad \alpha \in \R,
    \end{equation}
  where the normalization constant and the function $\rho(n)$ are given by
 \begin{equation}\label{normGK}
    \N(|z|^2)=\sum_{n=0}^{\infty}\frac {|z|^{2n}}{\rho(n)}, \qquad \rho(n)=[e_n]!.
 \end{equation}
   Briefly speaking,
  they replaced $-\infty < \gamma < \infty$ and $J>0$ in (\ref{GKCS})
   by $\alpha \in \R$ and $z \in \C$, respectively.
   We must emphasize the main difference between the
   GKCSs presented in (\ref {GKED}) and GK states in (\ref
   {GKCS}) in view of the significance and the role of $\gamma$ and
   $\alpha$, particularly in the integration procedure, in order to
   establish the resolution of unity.
   For this purpose, it is required to find an appropriate
   positive measure $d \lambda(z)$ such that
 \begin{equation}\label{RI-GKED}
     \int_0^{R}|z, \alpha\rangle \langle z, \alpha|d \lambda(z)=
     \sum_{n=0}^\infty|n\rangle\langle n|= \hat{I},
     \qquad 0 < R \leq \infty.
 \end{equation}
   Inserting (\ref {GKED}) in (\ref{RI-GKED}), writing $z= x e^{i\theta}$ and
   expressing the measure as
 \begin{equation}\label{measure}
   d \lambda(z)=d \lambda(|z|^2)=\pi \N(x^2) \sigma(x^2) x dx d \theta,
 \end{equation}
  performing the integration over $\theta \in [0, 2\pi]$, the over-completeness
  relation (\ref {RI-GKED}) finally
  reduced to the following moment problem (see \cite{Klauderetal2001} and Refs. therein)
 \begin{equation}\label{mom}
    \int_0 ^R  x^ n  \sigma(x) dx = \rho (n), \qquad 0 < R \leq \infty.
 \end{equation}

\subsection{\bf A discussion about the modification of GK states}\label{sec-Discusson}
    As it is observed in the previous subsection, in the
    modification imposed by El Kinani and Daoud on the GKCSs, the parameter $\alpha$
    has been {\it implicitly} considered as a constant,
    whose presence in the exponential factor
    of the introduced CSs preserves the temporal stability requirement
    (it is not now an integration {\it variable}).
    Meanwhile, for the
    temporal stability of the states in (\ref {GKED}) one reads:
 \begin{equation}\label{TS}
    e^{-i\hat{H}t}|z, \alpha\rangle = |z, \alpha'\rangle, \qquad
    \alpha'=\alpha + \omega t.
 \end{equation}
    Upon a closer inspection, one can see that the latter relation is indeed
    inconsistent with the
    resolution of the identity. By this, we mean that when $\alpha$
    is considered as a constant parameter, it really labels any  over-complete set of GKCSs,
    $\{ |z, \alpha\rangle \}$. But the time evolution
    operator in (\ref{TS}) maps the over-complete set of states $\{ |z, \alpha\rangle \}$
    to another over-complete set $\{ |z, \alpha'\rangle \}$.
    These are two {\it distinct set of CSs}, each of them labeled with a
    specific $\alpha$, if one consider the El kinani-Daoud formalism.
    But the temporal stability precisely means that under the choosen dynamics, the  time
    evolution of a CS remains CS, {\it of the same
    family}. In this manner, the states
    introduced in (\ref{GKED}) are not of the Gazeau-Klauder type, exactly.

    To overcome this problem,
    we redefine the resolution of the identity as follows:
 \begin{equation}\label{RI-RT}
  \lim_{\Gamma\rightarrow
   \infty}\frac{1}{2\Gamma}\int_{-\Gamma}^{\Gamma} d\alpha \int_0^R
    |z, \alpha\rangle \langle z,
     \alpha|d\lambda(z)=\sum_{n=0}^\infty|n\rangle\langle n|=
     \hat{I},   \qquad 0 < R \leq \infty.
 \end{equation}
    We can simplify the LHS of
    (\ref{RI-RT}) which interestingly led us exactly to the LHS of
    (\ref{RI-GKED}). Indeed we have
\begin{equation}\label{}
  \int_0^{R}|z, \alpha\rangle \langle z, \alpha|d \lambda(z)=
   \lim_{\Gamma\rightarrow
     \infty}\frac{1}{2\Gamma}\int_{-\Gamma}^{\Gamma} d\alpha \int_0^R
      |z, \alpha\rangle \langle z, \alpha|d\lambda(z),
 \end{equation}
   where $d\lambda(z)$ is determined as in (\ref{measure}).

  By this fact we want to conclude that both of
  the over-complete collection of states
  $\{|z, \alpha\rangle \}$ and $\{|z, \alpha'\rangle\}$, with {\it fixed} $\alpha$ and
  $\alpha'\equiv\alpha +\omega t$,
  belong to a large set of over-complete states
  with an arbitrary $\alpha$
\begin{equation}\label{largeH}
  \{|z, \alpha\rangle, z\in C, -\infty \leq \alpha \leq \infty\}.
\end{equation}
  Note that by replacing $\alpha \in R$ with $-\infty \leq \alpha \leq
  \infty$ in (\ref{largeH}) we want to emphasize that the parameter $\alpha$
  is relaxed from the constraint of being fixed.
  But we will encounter other difficulty, that is  the {\it variability} of $\alpha$ destroys the well definition of
  the {\it "operator valued function" $f(\alpha, \hat{n})$}, which will be
  introduced later in (\ref {nlGKED}),
  in addition to the deformed annihilation and creation operators $A=a f(\alpha, \hat n)$
  and $A^\dag  = f^\dagger(\alpha, \hat n)a^\dagger$.
  To overcome this difficulty we may bridge the gap between these two situations:
  variability and constancy of $\alpha$.
  We define the set of operators $A=a f(\alpha, \hat n)$, $A^\dag=f^\dag(\alpha,
   \hat n)a^\dagger$ and any other
  operator which explicitly depends on $\alpha$,
  in each {\it sector} ({\it subspace}) $\HH_\alpha$, labeled by a specific $\alpha$
  parameter, of the whole Hilbert space $\HH$
  which contains all GKCSs $\{|z, \alpha \rangle\}$.
  Indeed, the whole Hilbert space foliates by each
  $\alpha$ (remember the continuity of $\alpha$).
   Moreover, the action of the time evolution
   operator on any state on a specific sector, transfer it to another
   sector, both belong to a large Hilbert space.
   So, {\it when we deal with the operators that depend on the $\alpha$ parameter,
   it  should necessarily be fixed,
   while this is not the case when we are dealing with the states.}

  We notify here that taking $\alpha$ as a {\it constant} in somewhere and
  as a {\it variable} in another may be confused one and seems to be
  problematic. However, it is similar to the case which one
  encounters in the contexts of general relativity and quantum field theory,
  where the covariant formulation of the theory is required.
  In these cases one considers in the whole space with the dimension $n$, the space-like Cauchy hyper-surface
  $\Sigma$ is defined with the dimension $n-1$.
  Fixing $\alpha$ is similar to the so-called gauge fixing (e.g. a section in time).
  For instance by gauge fixing one may calculate the evolution of metric
 in solving the {\it Einstein equation} in general relativity or
 the commutation relation in quantum field theory.
 So in the present case, although the operators  are
 typically  true operators over the whole Hilbert space $\HH$, but the calculations and their commutation relations
 are done with a fixed $\alpha$ in  the subspace $\HH_\alpha$.


\subsection{\bf The Relation Between Nonlinear CS and GKCS}\label{sec-GKCS-NL}
  On the other hand the notion of nonlinear CSs introduced in
  \cite{Shanta1994,Matos1996,Man'ko1997} has attracted
  much attention in recent decade, especially in quantum optics.
  The realization of a special class of these states has been
  proposed in the quantized motion of a trapped ion in a Paul
  trap \cite{Matos1996, Vogel2002}.
  The nonlinear CS defined as eigenvector of the deformed
  annihilation operator has the following expansion over the Fock space
 \begin{equation}\label{nonl-cs}
    |z\rangle_{_f}=\N_f(|z|^2)^{-1/2}\sum_{n=0}^{\infty}\frac{z^n}
     {\sqrt{n!}[f(n)]!}|n\rangle, \quad [f(n)]!\doteq
    f(n)f(n-1)f(n-2)...f(1),
 \end{equation}
  where $ f(0)\equiv 1$ and $\N_f$ is some appropriate normalization constant
  may be determined.
  We recall that by replacing $f(n)$ with $\frac{1}{f(n)}$ in the
  relation (\ref {nonl-cs}) one
  immediately gets the nonlinear CSs  introduced in \cite{Roy-Roy2000}.
  We have called these states as the dual family of nonlinear CSs of
  Man'ko's type \cite{Ali2004}.

   Following the formalism given in \cite{Roknizadeh2004} for the states expressed
   in (\ref {GKED}) one may obtain
  \begin{equation}\label{nlGKED}
    f_{_{GK}}(\alpha, \hat{n})=e^{i\alpha(\hat{e}_n-\hat{e}_{n-1})}
    \sqrt{\frac{\rho(\hat{n})}{\hat{n}\rho(\hat{n}-1)}},
    \qquad \alpha, \quad \text{being fixed}.
  \end{equation}
    where we have choosed the notation $\hat{e}_n \equiv
    \rho(\hat{n})/\rho(\hat{n}-1)$.

   Moreover, we gain the opportunity to find raising and lowering operators
   in a safe manner
\begin{equation}\label{crea-anni}
  A_{GK}=a f_{GK}(\alpha, \hat{n}),  \qquad A^\dag_{GK}=
  f^\dag_{GK}(\alpha, \hat{n})a^\dag.
\end{equation}
    It is easy to verify that $A_{GK}|z, \alpha\rangle = z |z,
    \alpha\rangle$.
    Obviously the commutation relation between these two
   ($f-$deformed) ladder operators  obeys \cite{Man'ko1997} the relation
\begin{eqnarray}\label{com}
  [A_{GK}, A_{GK}^\dag] & = &\frac {\rho(\hat{n}+1)}{\rho(\hat{n})} -
  \frac {\rho(\hat{n})}{\rho(\hat{n}-1)}\nonumber\\
  &=& \hat{e}_{n+1}-\hat{e}_n.
\end{eqnarray}
   The special case $\rho(n)=n!$ recovers the standard bosonic
   commutation relation $[a, a^\dag]=\hat{I}$.
   Using the {\it "normal-ordered"} form of the Hamiltonian as in
    \cite{Roknizadeh2004} and taking $\hbar=1=\omega$,
     for the Hamiltonian of GKCSs we get
  \begin{equation}\label{normalH}
   \hat{H}_{_{GK}}\equiv \HD = A_{GK}^\dag A_{GK} = \hat{n}
   \Big|f_{_{GK}}(\alpha, \hat{n})\Big|^2 =
   \frac{\rho(\hat{n})}{\rho(\hat{n}-1)}=\hat{e_n}
  \end{equation}
   which clearly shows that the dynamics of the system is independent of $\alpha$.
 \section{{\bf The Dual family of GKCS as the temporally stable CS
          of the dual of KPS CS}}\label{sec-GK-temp}
  The KPS coherent states, introduced by Klauder, Penson and Sixdeniers
  \cite{Klauderetal2001} have the following form
 \begin{equation}\label{kps}
    |z \rangle_{KPS}=\N_{KPS}(|z|^2)^{-1/2}
    \sum_{n=0}^{\infty}\frac{z^n}{\sqrt{
    \rho(n)}}|n\rangle,  \qquad  z \in \C.
 \end{equation}
   As demonstrated in \cite{Roknizadeh2004}, all of the
   various sets of CSs introduced in \cite{Klauderetal2001}, constructed
   by diverse $\rho(n)$'s, are nonlinear CSs in nature. Also the appropriate
   nonlinearity function $f(n)$ as well as the deformed annihilation,
   creation and Hamiltonian operators were introduced there.
   Especially, it is found that $\rho(n)$ in (\ref{kps}) must satisfy the relation
   $\rho(n)=[nf^2(n)]!=[e_n]!$, where $e_n$'s are the eigenvalues of the associated
   factorized Hamiltonian.
   Taking into account the above results, comparing (\ref{kps}) with GKCS in
   (\ref{GKED}) and keeping $\hbar$ and  $\omega$ in the formulas, one may conclude that
 \begin{equation}\label{transfer}
    e^{-i \frac{\alpha}{\hbar \omega} \HD} | z \rangle_{KPS} = | z, \alpha
    \rangle, \qquad   0 \neq \alpha\in \R,  \qquad z\in \C.
 \end{equation}
   While $| z \rangle_{KPS}$ states are not temporally  stable, $|z, \alpha
   \rangle$ states enjoy this property.

   Now we may outline a relatively evident physical meaning to the arbitrary
   real $\alpha$ in (\ref{GKED}) or (\ref{transfer}) as
   $\alpha \equiv \omega t$,
   where by $t$ we mean the time that the operator acts on the
   KPS coherent states. It should be mentioned that, in a sense this interpretation has
   been presented for the GKCSs in a compact form \cite{Antoine2001}.
   If so, then $|z, \alpha \rangle$ can be considered as
   the evolution of $|z \rangle_{KPS}$.
   Therefore in a more general framework, we claim that the
   action of the evolution type operator
\begin{equation}\label{evolution}
  \hat{S}(\alpha) = e^{-i \frac{\alpha}{\hbar \omega} \HD},
  \qquad \hat S \hat S^\dag = \hat S^\dag \hat S = \hat{I}, \qquad 0 \neq \alpha \in
  \R,
 \end{equation}
   on any non-temporally stable CSs, makes it temporally
   stable CSs. So, {\it $\hat{S}(\alpha)$ is a
   nice and novel operator which transfer any generalized CS to a
   situation which it restores the temporal stability property.}
   Here, we stress on the fact that in (\ref {evolution}) the
    Hamiltonian, $\HD$ should satisfy
   $\HD|n \rangle = \hbar \omega e_n |n \rangle$.

    At this point we are ready to find a suitable way to define the dual family
    of GKCS. First, we note that the dual family of KPS CSs introduced in
    (\ref{kps}) has already been established in Ref. \cite{Roknizadeh2004},
    via the following exact form
\begin{equation}\label{kps-dual}
    |\widetilde z \rangle_{KPS}=\widetilde{\N}_{KPS}(|z|^2)^{-1/2}
    \sum_{n=0}^{\infty}\frac{z^n}{\sqrt{
    \mu(n)}}|n\rangle, \qquad z \in \C,
\end{equation}
    where
    \begin{equation}\label{relate-dual}
    \mu(n)\equiv \widetilde{\rho}(n)=\frac{(n!)^2}{\rho(n)}.
    \end{equation}
    Hereafter, the sign "tilde" over the operators and states, assign
    them to the corresponding dual operators and states, respectively.
    For instance $\widetilde{\rho}(n)$ is dual correspondence of
    $\rho(n)$.
    Eq. (\ref {relate-dual}) expresses the relation between
    KPS and the associated dual CSs, in a simple way.
    Obviously $\N_{KPS}$ and $\widetilde{\N}_{KPS}$
    in (\ref {kps}) and (\ref {kps-dual}) are the normalization
    constants may be obtained.
    Employing the formalism led to (\ref {transfer}),
    for "the dual of KPS states" in (\ref {kps-dual})
    naturally results in the following
    superposition of Fock states for the dual family of GKCSs
    (we shall refer to them as DGKCS)
\begin{eqnarray}\label{DGKED}
   \widetilde{\hat{S}}(\alpha) |\widetilde{z} \rangle_{KPS}  &=&
    e^{-i \frac{\alpha}{\hbar \omega}
    \widetilde{\hat{H}}}|\widetilde{z} \rangle_{KPS}
     = \widetilde{\N}(|z|^2)^{-1/2}
    \sum_{n=0}^{\infty}\frac{z^n e^{-i \alpha  \varepsilon_n}} {\sqrt{
    \mu(n)}}|n\rangle \nonumber\\ &=& |\widetilde{z, \alpha}
    \rangle,   \qquad z \in \C, \qquad 0 \neq \alpha \in \R,
\end{eqnarray}
    where $\widetilde{\N}=\widetilde{\N}_{KPS}$
     (because of the unitarity of $\hat S(\alpha)$, which preserves the norm)
     is given by
 \begin{equation}\label{normD}
     \widetilde{\N}(|z|^2)=\sum_{n=0}^{\infty}\frac{|z|^{2n}}{\mu(n)}.
 \end{equation}

     The special case of $\varepsilon_n=n$ in (\ref {DGKED}) recovers the canonical CSs,
     correctly.
     Note also that setting $\alpha=\omega t$ in (\ref{evolution})
     and (\ref{DGKED}) reduces the operators $\hat{S}(\alpha)$
      and $\widetilde{\hat{S}}(\alpha)$ to the well-known {\it time evolution
     operators} $\mathcal{U}(t)$ and
     $\widetilde{\mathcal{U}}(t)$, respectively.
     The case $\alpha = 0$ for the states in (\ref {GKED}) and (\ref {DGKED})
     recovers KPS and the corresponding dual CSs
     (which certainly are not temporally stable),
     respectively. The overlap between two states of the DGKCSs
     takes the following form
\begin{equation}\label{overlap}
     \widetilde{ \langle z, \alpha }| \widetilde{z',
     \alpha^ \prime}\rangle =
      \widetilde{\N}(|z|^2)^{-1/2} \widetilde{\N}(|z'|^2)^{-1/2}
      \sum_{n=0}^{\infty} \frac{(z^* z')^n e^{-i \varepsilon_n (-\alpha
      +\alpha')}}{\mu(n)},
\end{equation}
     which means that the states are essentially nonorthogonal.

    It should be noticed that the produced states
     ($\widetilde{|z, \alpha\rangle}$ introduced
    in (\ref {DGKED}))  form a {\it new class of
    generalized CSs} , essentially different from $|z, \alpha\rangle$ in (\ref {GKED}).
    Also, it is apparent that for our introduction of DGKCSs, we have obtained directly
    the analytic representation of DGKCS for any arbitrary quantum mechanical system.
    Using the formalism proposed in \cite{Roknizadeh2004},
    one can deduce the nonlinearity function for the dual states in (\ref {DGKED}) as
\begin{equation}\label{nlDGKED}
    \widetilde{f}_{GK}(\alpha, \hat{n}) = e^{i\alpha(\hat{\varepsilon}_n -
    \hat{\varepsilon}_{n-1})}\sqrt{\frac{\mu(\hat{n})}
    {\hat{n}\mu(\hat{n}-1)}},\qquad  \alpha \quad\text{being fixed},
  \end{equation}
    where we have used the notation
    $\hat{\varepsilon}_n \equiv\mu(\hat{n})/\mu(\hat{n}-1)$.
    Therefore, analogous to (\ref {normalH}) the deformed annihilation and creation operators
    of the dual system may be expressed explicitly as
 \begin{equation}\label{annihil-DGKED}
     \widetilde{A}_{GK}= a e^{i\alpha(\hat{\varepsilon}_n -
    \hat{\varepsilon}_{n-1})}\sqrt{\frac {\mu(\hat{n})}
    {\hat{n}\mu(\hat{n}-1)}},
 \end{equation}
 \begin{equation}\label{ceat-DGKED}
    \widetilde{A}^\dag _{GK}= e^{-i\alpha(\hat{\varepsilon}_n -
    \hat{\varepsilon}_{n-1})}\sqrt{\frac {\mu(\hat{n})}
    {\hat{n}\mu(\hat{n}-1)}} a ^\dag.
 \end{equation}
   The normal-ordered Hamiltonian of {\it dual oscillator} in the same manner
   stated in (\ref{normalH}) is
 \begin{equation}\label{Hamilt1}
   \widetilde{\HD}_{GK} \equiv \widetilde{\HD}=\widetilde{A}^\dag _{GK}\widetilde{A} _{GK} =  \frac{\mu(\hat{n})}
   {\mu(\hat{n}-1)} = \frac{\hat{n}^2}{\hat{e}_n},
 \end{equation}
    which is again independent of $\alpha$.
    As a result
 \begin{equation}\label{Hamilt2}
   \widetilde{\HD} |n\rangle =
    \widetilde{\mathcal{E}}_n |n\rangle \equiv \hbar\omega\varepsilon _n |n\rangle
    =\varepsilon _n |n\rangle,
    \qquad  \varepsilon_n \equiv \widetilde{e}_n=\frac{n^2}{e_n},
 \end{equation}
   where again we have used the units $\omega=1=\hbar$.
   The first equation in (\ref {Hamilt2}) illustrates clearly the relation between
   the eigenvalues of the two mutual dual systems.
   Also, the DGKCSs are  required to satisfy
   the following inequalities
 \begin{equation}\label{enIneq}
   0 = \varepsilon_0 < \varepsilon_1 <  \varepsilon_2< \cdots <
    \varepsilon_n < \varepsilon_{n+1} < \cdots,
 \end{equation}
    the same as that of $e_n$'s in (\ref {Ineq}).
    At this point a question may be arisen: to what extent one may be sure that the
    DGKCSs in (\ref {DGKED}) are of the Gazeau-Klauder  type? Let us briefly investigate this question.
  \begin{enumerate}
   \item
    {\it Continuity of labeling:} it is clearly satisfied.
\item
    {\it Resolution of unity:}
   \begin{equation}\label{RI-DGKED}
     \int_0^{\widetilde{R}}|\widetilde{z, \alpha}\rangle
     \langle\widetilde{z, \alpha}| d \lambda(z)
      = \sum_{n=0}^\infty|n\rangle\langle n|=\hat{I},
     \qquad 0 < \widetilde{R} \leq \infty,
   \end{equation}
    where the measure $d \lambda(z)$ is defined as in (\ref
    {measure}).
    The radius of convergence of DGKCS is determined as $\widetilde{R} = \lim _{n\rightarrow
    \infty}\sqrt[n]{\mu(n)}$ and
    $\mu(n)$ is defined as positive constants
    assumed to be appeared as moments of a probability distribution.
    Similar calculations led to the result in (\ref
    {mom}) are needed to arrive at the new moment problem associated with the
    DGKCSs
 \begin{equation}\label{momDual}
   \mu(n) \equiv \widetilde{\rho}(n)= \int_0^{\widetilde{R}}  x^n  \tilde{\sigma}(x)
   dx,   \qquad  0 < {\widetilde{R}}\leq \infty,
 \end{equation}
      which must be solved with the help of the early mentioned
      techniques.
      As for the GKCSs, we assume that $\mu(0)=1$ and $\mu(n) < \infty$ for all $n$.\\
 \item
  {\it Temporal stability:}
   using (\ref {DGKED})
   and the relevant Hamiltonian (\ref{Hamilt2}) gives us readily:
 \begin{eqnarray}\label{temp_dual}
     e^{-i \widetilde{\HD} t} |\widetilde{z, \alpha} \rangle &=&
     \widetilde{\N}(|z|^2)^{-1/2}\sum_{n=0}^{\infty}\frac{z^n e^{-i\varepsilon
     _n(\alpha+\omega t)}}{\sqrt{\mu(n)}}|n\rangle \nonumber \\
     & = & |\widetilde{z, \alpha+\omega t}\rangle,
  \end{eqnarray}
     which illustrates that DGKCSs remain coherent, as time goes on.
\item
   {\it Action identity:}
     from the condition (iii) we find that the time
     evolution of a CS is a map given by $(z, \alpha ) \mapsto (z, \alpha +\omega
     t)$. The new states $|\widetilde{z, \alpha }\rangle$  satisfy the relation:
 \begin{equation}\label{AI}
   \langle \widetilde{z, \alpha}| \widetilde{\HD} |\widetilde{z,
    \alpha}\rangle=\omega |z|^2,
 \end{equation}
   in consistence with Gazeau-Klauder's criteria,
   which is the so-called action identity. This is a strong
   requirement which uniquely specifies the weights $\{\mu(n)\}_{n \geq 0}$ in
   the denominator of the expansion coefficients of the  DGKCS.
   Using (\ref {DGKED}), (\ref {normD}) and (\ref {Hamilt2})
   in the LHS of (\ref {AI}) we obtain
  \begin{equation}\label{}
      \sum_{n=0}^\infty \frac{ \varepsilon_n |z|^{2n}}{\mu(n)}=|z|^2 \sum_{n=0}^\infty
      \frac{|z|^{2n}}{\mu(n)},
  \end{equation}
      by which we arrive at the following condition
  \begin{equation}\label{}
      \varepsilon_n = \frac{\mu(n)}{\mu(n-1)}.
   \end{equation}
   By conventional choice of $\mu(0) \equiv 1$, we thus deduce
   \begin{equation}\label{mudual}
    \mu(n)= \varepsilon_n\varepsilon_{n-1}...\varepsilon_1
    =\Pi_{k=1}^{n}\varepsilon_k\equiv[ \varepsilon_n]!.
  \end{equation}
\end{enumerate}
     So, we have established that the DGKCSs in (\ref {DGKED}) are
     exactly of the Gazeau-Klauder type.
    It ought to be noted that the same arguments we presented in section
    \ref{sec-Discusson} about the resolution of the identity
    (and the integration procedures),
    the $\alpha$ parameter (the states and the operators which depend on it)
    and the corresponding Hilbert spaces must also be considered
    for the DGKCSs have been built in the present section.

 \subsection{{\bf The introduction of temporally stable
               nonlinear CS}}\label{sec-NL-temp}
    Let us now outline the main idea in a general framework.
    It is believed that the property of the temporal stability is
    intrinsic to the {\it harmonic oscillator} and the systems which are
    {\it unitarily equivalent} to it \cite{Klauder2004}.
    But in what follows we shall demonstrate that how this
    important property can be restored by a redefinition of any generalized CSs
    which can be classified in the nonlinear CSs category.
    Recall that the nonlinear CSs  we introduced
    in (\ref {nonl-cs}) do not generally have  the temporal stability
    property \cite{Man'ko1997}.
    So, by considering the results obtained in the previous
    work \cite{Roknizadeh2004} and the above explanations, we want to go proceed
    and introduce generally the new notion of
    {\it "temporally stable nonlinear CSs"} as
 \begin{equation}\label{nonl-temp-cs}
  |z, \alpha
  \rangle_{f}=\N_f(|z|^2)^{-1/2}\sum_{n=0}^{\infty}\frac{z^n e^{-i
  \alpha e_n}}{\sqrt{n!}[f(n)]!}|n\rangle,
  \qquad e_n=nf^2(n), \qquad 0 \neq \alpha \in \mathbb{R}, \quad z\in
  \mathbb{C}.
 \end{equation}
    We can also define the dual of the latter states  by the following expression
 \begin{equation}\label{Roy-nonl-temp-cs}
     |\widetilde{z, \alpha}
     \rangle_{f}=\widetilde{\N}_f(|z|^2)^{-1/2}
     \sum_{n=0}^{\infty}\frac{z^n [f(n)]! e^{-i \alpha  \varepsilon_n}}
     {\sqrt n!} |n \rangle, \quad  \varepsilon_n
     = \frac{n}{f^2(n)}, \quad 0 \neq \alpha \in \mathbb{R}, \quad z\in
     \mathbb{C},
 \end{equation}
    which are indeed the temporally stable version of the nonlinear CSs
    have been introduced in \cite{Roy-Roy2000}.
    In both sets of the CSs given by (\ref {nonl-temp-cs}) and (\ref
   {Roy-nonl-temp-cs}), $\alpha$ is a real constant and the normalization
   factors are independent of $\alpha$.
   Setting $\alpha = 0$ in (\ref {nonl-temp-cs}) and (\ref
   {Roy-nonl-temp-cs}), we recover the old form of Man'ko's and
   Roy's nonlinear CSs, respectively, which clearly are not temporally stable.

 \subsection{\bf Temporally stable CS of $SU(1, 1)$ group}

   An instructive example of the families of nonlinear CSs
   is provided by the Gilmore-Perelomov(GP) \cite{Perelomov} and
   Barut-Girardello(BG) CSs \cite{Barut1971}, defined for the
   discrete series representations of the group $SU(1,1)$.
   Using the results of \cite{Roknizadeh2004} for GP states,
   and then imposing the proposed formalism on them, the
   {\it "temporally stable CSs of GP type associated with $SU(1, 1)$ group"}
   can be defined as
 \begin{equation}\label{gilperCS}
    |z, \alpha\rangle_{GP}^{SU(1, 1)} = \N_{GP}(|z|^2)^{-1/2}\sum_{n=0}^\infty
    \frac{z^n e^{-i \alpha \frac {n}{(n+2\kappa-1)}}}
    {[n!/\Gamma(n+2\kappa)]^{1/2}}|n\rangle,    \qquad |z|<1,
 \end{equation}
  where $\N_{GP}$ is a normalization factor
  and the parameter $\kappa
  =1,3/2,2, 5/2, \cdots$, labels the $SU(1,1)$ representation being
  used.
  Analogously, applying the presented extension to the BG type of CSs, gives immediately
  {\it "temporally stable CSs of BG type associated with $SU(1, 1)$
  group"} as follows
 \begin{equation}\label{bargirCS}
   |z, \alpha\rangle_{BG}^{SU(1, 1)} \equiv  |\widetilde{z, \alpha}\rangle_{GP}^{SU(1, 1)}=
   \N_{BG}(|z|^2)^{-1/2}
   \sum_{n=0}^\infty\frac{z^n e^{-i \alpha n(n+2\kappa-1)}}{[n!\Gamma(n+2\kappa
   )]^{1/2}} |n\rangle,    \qquad  z \in \C,
 \end{equation}
  where once more, $\N_{BG}$ is chosen by normalization of the states.


 \subsection{\bf Temporally stable CS of Penson-Solomon type and its dual}
    As established in \cite{Roknizadeh2004}, the generalized CSs
    introduced by Penson and Solomon \cite{Penson1999} as
 \begin{equation}\label{ps}
   |q, z\rangle = \N(q, |z|^2)\sum _{n=0}^\infty \frac  {q^{\frac{n(n-1)}{2}}}
   {\sqrt{n!}}z^n |n \rangle,
 \end{equation}
  are also nonlinear with $f(n)= q^{(1-n)}$ and therefore the
  factorized Hamiltonian reads $\HD_{PS}=\hat{n}q ^{2(1- \hat{n})}$.
  It is stated in \cite{Penson1999} that under the
  action of $\exp(-i \hat{H}t)$ these
  states are temporally stable, where the Hamiltonian $\hat{H}=a^\dag a=\hat{n}$
  expresses the (shifted) quantum harmonic oscillator
  with the corresponding canonical CS.
  Seemingly to verify the invariance under time evolution operator,
  it may be more realistic to act the operator, $\exp(-i \HD_{PS}t)$,  on
  the PS states in (\ref{ps}),
  where $\HD_{PS}|n\rangle=\hat nq ^{2(1-\hat n)}|n\rangle=nq ^{2(1-n)}|n\rangle$.
  Clearly by such proposition
  these states are not temporally stable.
  But  the presented formalism in this paper allows one to
  construct the temporally stable
  CS of PS type as follows
\begin{equation}\label{ps-temp}
  |q, z, \alpha \rangle \equiv e^{-i \frac{\alpha}{\hbar \omega} \HD_{PS}}
   |q, z\rangle = \N_{PS}(q, |z|^2)\sum _{n=0}^\infty \frac  {q^{\frac{n(n-1)}{2}}}{\sqrt{n!}}
    e^{-i \alpha e_n}  z^n |n \rangle,
 \end{equation}
  where $e_n=nq ^{2(1-n)}$.

  We have already introduced the dual family of the PS states of Eq.
  (\ref{ps}) $\widetilde{|q, z\rangle}$ in Ref. \cite{Roknizadeh2004}.
  So the temporally stable dual of these states may also be
  obtained immediately as
\begin{equation}\label{ps-temp}
   |\widetilde{q, z, \alpha} \rangle \equiv e^{-i \frac{\alpha}{\hbar \omega}
    \widetilde{\HD}_{PS}}
    \widetilde{|q, z\rangle} = \widetilde{\N}_{PS}(q, |z|^2)\sum _{n=0}^\infty \frac
    {q^{\frac{-n(n-1)}{2}}}{\sqrt{n!}}
     e^{-i \alpha \varepsilon_n}  z^n |n \rangle,
 \end{equation}
 where $\varepsilon_n=\frac{n}{q^{2(1-n)}}$.


 \subsection{\bf Some remarkable points}
    We end this section with some remarkable points.
 \begin{itemize}
    \item
    First, one can prove that the
    $f$-deformed annihilation operator
    given by (\ref {nlGKED}) is just the same as the one
    derived earlier in \cite{Antoine2001}, denoted by $a(\alpha)$
  \begin{equation}\label{anni-Antoin}
    a(\alpha)=e^{-i \alpha \hat{H}(\hat{n})/\hbar\omega} \check{a}
    e^{i \alpha \hat{H}(\hat{n})/\hbar\omega},
  \end{equation}
    or in terms of the introduced evolution operator
    $\hat{S}(\alpha)$,
    \begin{equation}\label{anni-Antoine}
    a(\alpha)= \hat{S}(\alpha) \check{a} \hat{S}^\dag(\alpha).
  \end{equation}
     It must be noticed that $\check{a}$ and its adjoint $\check{a}^\dag$
     in Eqs. (\ref {anni-Antoin})
     and (\ref{anni-Antoine}),
     have been defined as follows
\begin{equation}\label{anni}
    \check{a}|n\rangle = \sqrt{e_n} |n-1\rangle, \qquad
    \check{a}^\dag |n\rangle = \sqrt{e_{n+1}} |n+1\rangle.
\end{equation}
     Now using the relations $f^\dag(\hat{n}) a = a
     f^\dag(\hat{n}-1)$ and $\hat{n}a =a (\hat{n}-1)$, we have
  \begin{equation}\label{anniAntoin}
     e^{-i \alpha \hat{H}(\hat{n})/\hbar\omega} \check{a}
     = \check{a} e^{-i \alpha \hat{H}(\hat{n}-1)/\hbar\omega}=
       \check{a} e^{-i \alpha \hat{e}_{n-1}},
  \end{equation}
    where we have used Eq. (\ref {kps-14}) in the last step. Upon replacing
    (\ref {anniAntoin}) in the RHS of (\ref {anni-Antoin}) and taking into account
    (\ref {kps-14}) we are readily led to the equality
    $a(\alpha) \equiv A_{GK} = af_{GK}(\alpha, \hat{n}).$
\item
   In the light of the presented explanations
   the annihilation operator eigenstate (algebraic definition) for GKCSs and DGKCS
   are such that
 \begin{equation}\label{aniGKED-def}
     A_{GK}|z, \alpha \rangle = z |z, \alpha \rangle, \qquad
     \widetilde{A}_{GK} |\widetilde{z, \alpha }\rangle = z |\widetilde{z, \alpha} \rangle.
 \end{equation}
    The deformed annihilation and creation operators $\widetilde{A}_{GK}$
    and $\widetilde{A}^\dag_{GK}$ of the dual oscillator algebra,
    satisfy the following eigenvector equations
 \begin{equation}\label{GK-da1}
   \widetilde{A}_{GK}|n\rangle = \sqrt{ \varepsilon_n}
   e^{i\alpha ( \varepsilon_n-\varepsilon_{n-1})}|n-1\rangle
 \end{equation}
 \begin{equation}\label{GK-da2}
   \widetilde{A}^\dag_{GK} |n\rangle = \sqrt{\varepsilon_{n+1}}
   e^{i\alpha(\varepsilon_{n+1}- \varepsilon_n)}|n+1\rangle
\end{equation}
\begin{equation}\label{GK-da3}
   [\widetilde{A}_{GK}, \widetilde{A}^\dag_{GK}]|n\rangle =
   (\varepsilon_{n+1}- \varepsilon_{n})|n\rangle
 \end{equation}
 \begin{equation}\label{GK-da4}
   [\widetilde{A}_{GK}, \hat{n}] =
   \widetilde{A}_{GK} \qquad [\widetilde{A}^\dag_{GK}, \hat{n}] = -\widetilde{A}^\dag_{GK}.
 \end{equation}
   Upon looking on the actions defined in (\ref{GK-da1}) and (\ref{GK-da2})
   one can interpret $\widetilde{A}_{GK}$ and $\widetilde{A}^\dagger_{GK}$ as
   the operators which correctly annihilate and create
   one quanta of {\it deformed photon}, respectively.
   A closer look at the basis of the involved Hilbert space $\HH_\alpha$
   in each over-complete set $\{|z, \alpha\rangle\}$, shows that
   it is spanned by the vectors
 \begin{equation}\label{spann-H}
   |n, \alpha\rangle=\frac{(\widetilde{A}^\dag_{GK})^n e^{i \alpha
    \varepsilon_n }}{\sqrt {[e_n]!}} |0\rangle  \equiv |n \rangle,    \qquad
   \widetilde{A}_{GK} |0\rangle =0.
 \end{equation}
   Moreover, we have omitted the $\alpha$ parameter from the basis for simplicity.
   At last, we are able to introduce the generators of the deformed oscillator
   algebra \cite{Borzov1997} of Gazeau-Klauder and the corresponding dual family
   as $\{A_{GK}, A^\dag_{GK}, \HD \}$ and $\{\widetilde{A}_{GK}, \widetilde{A}^\dag_{GK},
   \widetilde{\HD}\}$, respectively.
 \item
   One may expect that the inequalities for $ \varepsilon_n$ given in
   (\ref {enIneq}) corresponding to any solvable system
   do not hold for DGKCSs (the restriction which also exists in (\ref {Ineq}) for GKCSs).
   This is generally may be true, but fortunately many cases-if not all-
   such as all physical systems will be considered in this paper are of this sort
   (both of $ \varepsilon_n$ and $e_n$ are strictly increasing). So it must
   be mentioned that before making use of our formalism for dual states associated
   with any set of GKCSs one should be sure about the
   condition (\ref {enIneq}). If both of the
   inequalities in (\ref {Ineq}) and (\ref {enIneq}) hold simultaneously, then one has:
\begin{equation}\label{Ineq-en}
  1>\frac{e_n}{e_{n+1}}>\frac{n^2}{(n+1)^2}
\end{equation}
   which can be expressed in terms of the nonlinearity function
    $f_{GK}(\alpha, \hat{n})$ as follows
\begin{equation}\label{Ineq-fn}
   \sqrt{\frac{n+1}{n}} > \sqrt{ \left|\frac{f_{GK}(\alpha, \hat{n})}
   {f_{GK}(\alpha, \hat{n}+1)}\right|^2 } > \sqrt{\frac{n}{n+1}},
\end{equation}
   for all $ n > 0 $.
 \item
   As it may be clear, when one wants to work with one of the dual
   pairs singly, they can be considered on their relevant
   domains. But to deal with their mutual relation, calculations must be done only
   in the intersection of the domains of the pair of
   CSs (in this case GKCSs and DGKCSs); i.e., generally on a unit
   disk, unless the CS is defined on a finite dimensional
   Hilbert space. As we shall see later, the latter is the case for Morse potential.
\item
   And finally, the probability distribution for
   the DGKCSs is defined as
\begin{equation}\label{distribut-Ev}
   \widetilde{\mathbf{P}}(n) = \left|\langle n |\widetilde{z,
   \alpha}\rangle\right|^2=\widetilde{\N}(|z|^2)^{-1}\frac{|z|^{2n}}{\mu(n)},
\end{equation}
   which is independent of $\alpha$ parameter.
 \end{itemize}

   We terminate this section with recalling that there exists
   also a set of equations such as (\ref
   {GK-da1}-\ref {GK-da4}) related to GKCSs which may be obtained just by replacing
   $\widetilde{A}_{GK}$, $\widetilde{A}^\dag_{GK}$ and $ \varepsilon_n$
   with $A_{GK}, A_{GK}^\dag$ and $e_n$,
   respectively. The latter have been already derived by applying SUSYQM
   techniques \cite{El Kinani2002}, but re-derivation of them are very easy by our
   formalism.
   According to their results, the one-dimensional SUSYQM provides
   a mathematical tool to define ladder operators for an exactly solvable potentials.
   But the authors did not express the {\it explicit} form of
   the ladder operators, and only the concerning actions were expressed there.
   Therefore besides the simplicity of our method, it is more complete
   in the sense that as we found the {\it explicit} form of the
   raising and lowering operators in terms of  the standard bosonic creation
   and annihilation operators and
   the photon number (intensity of the field) have been found
   easily (see equations (\ref {crea-anni}), (\ref{annihil-DGKED}), (\ref
   {ceat-DGKED})).


\section{\bf Displacement operators associated with GKCS and
    the corresponding dual family}\label{sec-displca}
    Now, which we introduced the explicit form of the
    deformed annihilation operator (and hence the annihilation operator definition
    for GKCSs and DGKCSs according to equations in
    (\ref {aniGKED-def}),
    we are in the position  to extract the CSs of Klauder-Perelomov type
    for an arbitrary quantum mechanical system.
    For this purpose, we introduce the following auxiliary operators related to GKCSs
 \begin{equation}\label{DisplaceGK}
   B_{GK} =a \frac{1}{f_{_{GK}}(-\alpha, \hat{n})},
   \qquad B_{GK}^\dag =  \frac{1}{f^\dag_{_{GK}}(-\alpha, \hat{n})}
   a^\dag,
 \end{equation}
   and those for the dual families DGKCSs
 \begin{equation}\label{DisplaceDGK}
   \widetilde{B}_{GK} =a \frac{1}{\widetilde{f}_{GK}(-\alpha, \hat{n})}, \qquad
   \widetilde{B}^\dag _{GK}=\frac{1}{ \widetilde{f}^\dagger_{GK}(-\alpha, \hat{n}) } a^\dag.
 \end{equation}
   Notice that the minus sign in the argument of the $f$-function
   is needed in both cases, since only in such cases we have
   $f^\dag_{GK}(-\alpha, \hat{n})=f_{GK}(\alpha, \hat{n})$.

   The $f$-deformed operators given by (\ref {DisplaceGK}) and (\ref{DisplaceDGK})
   are canonically conjugate of the $f$-deformed creation and annihilation
   operators  $(A_{GK}, A_{GK}^\dag)$ and $(\widetilde{A}_{GK},
   \widetilde{A}^\dag_{GK})$, respectively;
   i.e., they satisfy the algebras $[A_{GK}, B_{GK}^\dag]=[B_{GK},
   A_{GK}^\dag]=\hat{I}$ and
   $[\widetilde{A}_{GK}, \widetilde{B}^\dag_{GK}]=[\widetilde{B}_{GK},
   \widetilde{A}^\dag_{GK}]=\hat{I}$, respectively.
   Now we have all mathematical tools to construct the
   displacement operators for GKCS
\begin{equation}\label{disGKED}
   D_{GK}(z, \alpha)=\exp(z B_{GK}^\dag - z^* A_{GK}),
\end{equation}
   and for DGKCS,
\begin{equation}\label{disDGKED}
  \widetilde{D}_{GK}(z, \alpha) = \exp(z \widetilde{B}^\dag _{GK}- z^*
  \widetilde{A}_{GK}).
\end{equation}
   The actions of $D_{GK}(z, \alpha)$ and $\widetilde{D}_{GK}(z, \alpha)$
   on the vacuum state $| 0 \rangle$ yield the GKCSs and DGKCS, up
   to some normalization constant,respectively, as we demand.
   From the group theoretical point of view,
   one can see that the sets $\{A_{GK}, B _{GK}^\dag, B_{GK}^\dag A_{GK}, \hat I\}$ and
   $\{\widetilde{A}_{GK}, \widetilde{B}^\dag_{GK},
   \widetilde{B}^\dag_{GK} \widetilde{A}_{GK}, \hat I\}$,
   which are respectively associated with GKCSs and DGKCSs, form the
   Lie algebra $h_4$ and the corresponding Lie group is the
   well-known Weyl-Heisenberg(W-H) group. Also, the action of the latter operators
   in (\ref {disGKED}) and (\ref {disDGKED})
   on the vacuum state are the orbits of the projective {\it non-unitary}
   representations of the W-H group \cite{Ali2004}. It must be
   understood that as we pointed out earlier, we have applied neither the
   formalism in \cite{Ali2004} nor the equivalent formalism of
   Ref. \cite{Roy-Roy2000} for constructing the
   dual states, since the states obtained from the earlier formalisms
   were not full consistent with the Gazeau-Klauder's criteria.
   Indeed, we proposed a rather new way,
   through viewing the GKCSs $|z, \alpha\rangle$ and its dual pair
   $|\widetilde{z, \alpha}\rangle$ as
   generalization of KPS nonlinear CSs $|z\rangle_{KPS}$ and its dual
   $|\widetilde{z}\rangle_{KPS}$ to the two distinct temporally stable CSs, respectively.
   Speaking otherwise, the operators introduced in (\ref {disGKED}) and
   (\ref {disDGKED}) do not have the relation:
   $\widetilde{D}_{GK}(-z, \alpha)=D_{GK}(z, \alpha)=[D_{GK}(z,
   \alpha)^{-1}]^\dag$, which is the characteristic of the
   earlier formalisms.
   To this end, it is possible to build the following displacement type
   operators,
  $ V_{GK}(z, \alpha)=\exp(z A_{GK}^\dag -z^* B_{GK})$
     for GKCSs, and in a similar manner,
  $ \widetilde{V}_{GK}(z, \alpha)=\exp(z \widetilde{A}^\dag _{GK}- z^*
  \widetilde{B}_{GK})$
    for DGKCSs, whose actions on the vacuum state, yield two new sets of states.
    But it is easy to investigate that none of them
    can be classified in the Gazeau-Klauder CSs.


 \section{\bf The construction of even, odd and Schr\"{o}dinger cat
           coherent states from the introduced DGKCS}\label{sec-displca}
  Various superpositions of CSs may result in different nonclassical states of
  light. Recently, there has been much interest in the construction as well as generation
  of these states, in the regard of their applications in the context of
  quantum optics. Their different characteristics are due to the various
  quantum interference between summands.
  As an example, the even and odd CSs associated with canonical CSs as
  well as other classes of generalized CSs such as
  nonlinear CSs extensively studied in the literature \cite{Mancini1997}
  exhibit nonclassical features, such as squeezing, sub-Poissonian
  statistics (antibunching) and oscillatory number distribution.
  The symmetric (antisymmetric) combinations of GKCSs
  have been introduced in \cite{El Kinani2001}. Similarly using the unnormalized
  DGKCSs we are led to the even(odd) CSs denoted by $+(-)$:
 \begin{eqnarray}\label{even-odd}
   |\widetilde{z, \alpha} \rangle_\pm  &=&
   \widetilde{\N'}_\pm(|z|^2)^{-1/2}\left(|\widetilde{z, \alpha} \rangle \pm
   |\widetilde{-z, \alpha} \rangle\right)\nonumber \\&=&
   \widetilde{\N'}_\pm(|z|^2)^{-1/2}
   \sum_{n=0}^{\infty}\frac{e^{-i \alpha  \varepsilon_n }[z^n\pm(-z)^n]}{\sqrt{
   \mu(n)}}|n\rangle,
 \end{eqnarray}
   where $z\in \C$ and  $\alpha \in \R$.
   For the normalization factors we get
 \begin{equation}\label{norm-eo}
   \widetilde{\N'}_\pm(|z|^2) =2 \left(\sum_{n=0}^{\infty}\frac{|z|^{2n}}{\mu(n)} +
   \sum_{n=0}^{\infty}\frac{(-1)^n |z|^{2n}}{\mu(n)}\right)^{-1}.
 \end{equation}
    A few simplification imposed on the states in (\ref {even-odd}) will clarify the name
    even(odd) associated with these states
\begin{equation}\label{even2}
  |\widetilde{z, \alpha }\rangle_+ =
    \widetilde{\N'}_+(|z|^2)^{-1/2}
    \sum_{n=0}^{\infty}\frac{z^{2n}e^{-i \alpha \varepsilon_{_{2n}}}}{\sqrt{
    \mu(2n)}}|2n\rangle,
 \end{equation}
 \begin{equation}\label{odd2}
     | \widetilde{z, \alpha} \rangle_- =
    \widetilde{\N'}_-(|z|^2)^{-1/2}
    \sum_{n=0}^{\infty}\frac{z^{2n+1}e^{-i \alpha \varepsilon_{_{2n+1}}}}{\sqrt{
    \mu(2n+1)}}|2n+1\rangle.
 \end{equation}
    Finally the $(\pm)$ states in (\ref {even2}) and (\ref {odd2}) satisfy
   the following eigenvalue equations
 \begin{equation}\label{a2}
    (\widetilde{A}_{GK})^2 |\widetilde{z, \alpha} \rangle_\pm =
    z^2 |\widetilde{z, \alpha }\rangle_\pm.
 \end{equation}
    The probability distributions for the even-DGKCS $(+)$ and odd-DGKCS $(-)$ are derived as
 \begin{equation}\label{distribut-Ev}
   \widetilde{\mathbf{P}}_\pm(n) = \left|\langle n |\widetilde{z,
   \alpha }
   \rangle_\pm\right|^2=\widetilde{\N'}_\pm(|z|^2)^{-1}\frac{|z|^{2n}}{\mu(n)},
\end{equation}
   which clearly are independent of the $\alpha$ parameter.

   Now, we pay attention to another specific superposition of the DGKCSs
   $|\widetilde{z, \alpha} \rangle$, by which we may
   obtain the {\it real} $(+)$ and {\it imaginary} $(-)$ {\it Schr\"{o}dinger cat states} as
 \begin{equation}\label{cat1}
       |\widetilde{z, \alpha} \rangle_{\pm}^{Cat} =
       \widetilde{\N''}_\pm (|z|^2)^{-1/2} (|\widetilde{z, \alpha} \rangle \pm
       |\widetilde{z^*, \alpha} \rangle),
 \end{equation}
    where $z^*$ is the complex conjugate of $z$. Inserting $z=r e^{i \theta}$
    in the last equations give us the following explicit forms
 \begin{equation}\label{cat2}
    | \widetilde{z, \alpha} \rangle_+^{Cat} =
     \widetilde{\N''}_+(|z|^2)^{-1/2} \sum_{n=0}^\infty \frac{r^n \cos
     (n\theta)}{\sqrt{\mu(n)}}e^{-i \alpha  \varepsilon_n}|
     n\rangle
  \end{equation}
  and
 \begin{equation}\label{cat3}
    | \widetilde{z, \alpha} \rangle_-^{Cat} =
     \widetilde{\N''}_-(|z|^2)^{-1/2} \sum_{n=0}^\infty \frac{r^{n+1} \sin
     [(n+1)\theta]}{\sqrt{\mu(n+1)}}e^{-i \alpha \varepsilon_{n+1}}|
     n+1\rangle,
 \end{equation}
   where the normalization constants would be find as
 \begin{equation}\label{cat4}
     \widetilde{\N''}_+(|z|^2) =\sum_{n=0}^\infty \frac{r^{2n} \cos^2
     (n\theta)}{\mu(n)}
 \end{equation}
   and
 \begin{equation}\label{cat5}
   \widetilde{\N''}_-(|z|^2) =  \sum_{n=0}^\infty \frac{r^{2(n+1)} \sin ^2
   [(n+1)\theta]}{\mu(n+1)}.
 \end{equation}
   The probability distribution for the the real ($+$) and imaginary ($-$)
   Schr\"{o}dinger cat CSs in (\ref {cat2}) and (\ref {cat3}) can be calculated as
\begin{equation}\label{distribut-Ev}
   \widetilde{\mathbf{P}}_ \pm ^{Cat}(n) = \left|\langle n |\widetilde{z,
   \alpha} \rangle_\pm^{Cat}\right|^2 = \widetilde{\N''_\pm}(r^2)^{-1}
   \frac{r^{2n}(1 \pm \cos(2n \theta))}{\mu(n)},
\end{equation}
  which  are again independent of $\alpha$.


  \section{\bf Some physical applications of the DGKCS}\label{sec-Examples}
   In order to illustrate the presented idea in this paper,
   let us apply the formalism on some physical examples
   which the associated GKCSs have already been known.
   To economize in the space the complete form of DGKCSs have not
   been given in what follows and it will be enough
   for our intention to present $\varepsilon_n$,
   $\mu(n)$ and $\widetilde{\N}(|z|^2)$, since substituting these quantities into
   (\ref{DGKED}) gives readily the explicit form of the DGKCSs,
    $|\widetilde{z, \alpha }\rangle$.
   \vspace{4 mm}\\
  {\bf Example} $1$  {\bf Harmonic oscillator:}
\vspace{2 mm}\\
   As the simplest example we apply the formalism to the harmonic
   oscillator Hamiltonian, whose nonlinearity function is
   equal to $1$, hence $ \varepsilon_n = n=e_n$ which results in the moments
   as $\mu(n)=n!=\rho(n)$. Note that we have considered a shifted Hamiltonian to lower the
   zero-point energy to zero ($e_0=0=\varepsilon_0$). Eventually
\begin{equation}\label{}
    |\widetilde{z, \alpha }\rangle_{CCS}=e^{-|z|^2/2}
    \sum_{n=0}^{\infty}\frac{z^n e^{-i \alpha n}}{\sqrt{
    n!}}|n\rangle = |z, \alpha \rangle _{CCS}
\end{equation}
    ensures the {\it self-duality} of canonical CS.
    For this example all the Gazeau-Klauder's requirements are satisfied, trivially.
 \vspace{4 mm}\\
  {\bf Example} $2$  {\bf P\"{o}schl-Teller potential:}
\vspace{2 mm}\\
    The interest in this potential and its CSs is due to various applications in
    many fields of physics particularly in  atomic and molecular physics.
    The usual GKCSs for the
    P\"{o}schl-Teller potential have been demonstrated nicely by J-P Antoine
    {\it et al} \cite{Antoine2001}. Their obtained results are as follows
 \begin{equation}\label{poshCnd}
    e_n=n(n+\nu),   \qquad  \rho(n)=\frac{n!\Gamma(n+\nu+1)}{\Gamma(\nu+1)}, \qquad \nu > 2
 \end{equation}
   with the radius of convergence $R=\infty$.
   Consequently  using the Eqs. given by (\ref {poshCnd})
   in (\ref {relate-dual}) and (\ref{Hamilt2})
   we are able to construct DGKCSs associated with this particular potential by the new
   quantities obtained as
 \begin{equation}\label{poshCnd-Dual1}
    \varepsilon_n=\frac{n}{n+\nu}, \qquad \mu(n)=\frac{n!
    \Gamma(\nu+1)}{\Gamma(n+\nu+1)}, \qquad \nu > 2
 \end{equation}
     and for the normalization constant we obtain from (\ref{normD})
 \begin{equation}\label{poshCnd-Dual2}
     \widetilde{\N}(|z|^2)=(1-|z|^2)^{-1-\nu},
 \end{equation}
   whose region of convergence is determined as the open unit disk.
   The overlap between two of these states when $\alpha= \alpha'$ is obtained
   from (\ref{overlap}) as
\begin{equation}
    \langle \widetilde{ z, \alpha} |  \widetilde{z', \alpha }\rangle=
   \left[(1-|z|^2)(1-|z'|^2)\right]^{(1+\nu)/2}(1- z z')^{(-1-\nu)}.
\end{equation}
   To be ensure, for these dual states we only investigate the
   resolution of the identity; since the other three requirements
   are satisfied obviously. As required,  we have to find $\tilde{\sigma}(x)$
   such that the moment integral
\begin{equation}\label{poshCnd-Dua3}
    \int_0^1 x^n \tilde{\sigma}(x) dx=\frac{n! \Gamma(\nu+1)}{\Gamma(n+\nu+1)}
\end{equation}
    holds. It may be checked that the proper weight function $\tilde{\sigma}(x)$
    is determined as $\nu (1-x)^{\nu-1}$.

    At this point we recall that $e_n$'s in (\ref {poshCnd}) denotes
    the eigenvalues of different Hamiltonians.
    The characteristics of the dynamical system has been shown  by the parameter $\nu$.
    For instance, the eigenvalues of the anharmonic (nonlinear) oscillator,
    well studied in literature and the related
    GKCSs and GK states have been discussed
    in Refs. \cite{El Kinani2001} and \cite{Roy2003} in details, respectively.
    In current example the parameter $\nu$ is related to
    two other parameters, namely $\lambda$ and $\kappa$ through the relation $\nu=\lambda
    +\kappa$, which determine the height and the depth of the well
    potential. However, when one deals with the nonlinear oscillator it has
    another meaning; e.g. we refer to Ref. \cite{Roy2003},
    in which the interest was due to its usefulness in the study
    of laser light propagation in a {\it nonlinear Kerr medium}.
    In particular, $\nu$ in this case is related to the
    nonlinear susceptibility of the medium. So, the obtained results in (\ref
    {poshCnd-Dual1})-(\ref {poshCnd-Dual2}) can be exactly used for the
    anharmonic oscillator, too. To this end,  we shall see in the next example
    that the case of $\nu=2$ in (\ref {poshCnd}) is the
     eigenvalues of the infinite  well potential.
     \vspace{4 mm}\\
   {\bf Example} $3$  {\bf Infinite well potential:}
\vspace{2 mm}\\
    The GKCSs for the infinite
    well, have been established by J-P Antoine {\it et al} in \cite{Antoine2001}. The
    related quantities are
\begin{equation}\label{well}
   e_n=n(n+2), \qquad \rho(n)=\frac{n!(n+2)!}{2},
\end{equation}
   with radius of convergence as $R=\infty$.
   Consequently  inserting (\ref{well}) in (\ref {relate-dual}) and (\ref{Hamilt2})
   one can construct the dual of these states by the quantities
\begin{equation}\label{}
   \varepsilon_n=\frac{n}{n+2},  \qquad \mu(n)=\frac{2}{(n+1)(n+2)}
\end{equation}
   and the normalization factor can be obtained from (\ref {normD}) as
\begin{equation}\label{}
    \widetilde{\N} (|z|^2)=(1-|z|^2)^{-3}.
\end{equation}
   whose region of convergence is determined as the open unit disk.
   The overlap between two of these states for the special case  $\alpha= \alpha'$ is
   obtained from (\ref{overlap}) as
\begin{equation}
   \langle \widetilde{ z, \alpha} |  \widetilde{z', \alpha }\rangle=
    \left[ (1-|z|^2)(1-|z'|^2) \right]^{3/2}
   \frac{1}{1-z z'}.
\end{equation}
   To clarify the fact that these dual states are actually CSs, we only investigate the
   resolution of the identity, since  the other three requirements
   are satisfied straightforwardly. For this condition we have to find
   $\tilde{\sigma}(x)$ such that the integral
\begin{equation}\label{}
    \int_0^1 x^n \tilde{\sigma}(x) dx=\frac{2}{(n+1)(n+2)}
\end{equation}
    holds. It is easy to verify that $\tilde{\sigma}(x)=2(1-x)$ is the solution.
\vspace{4 mm}\\
    {\bf Example} $4$  {\bf Morse potential:}
\vspace{2 mm}\\
    The GKCSs for the Morse potential, which is the simplest type of anharmonic oscillator
    and is useful in various problems in
    different fields of physics(for example spectroscopy,
    diatomic and polyatomic molecule vibrations and scattering),
    can be obtained using the related quantities given in \cite{Popov2003}:
\begin{equation}\label{morse}
    e_n = \frac{n(M+1-n)}{M+2}, \qquad \rho(n)=\frac{\Gamma(n+1)\Gamma(M+1)}
    {(M+2)^n \Gamma(M+1-n)},
\end{equation}
    where $n=0,1,2, \cdots < (M+1)$.
    Therefore taking into account (\ref{morse}) in (\ref{relate-dual}) and (\ref{Hamilt2})
    the dual of these states
    can be produced by the following quantities
\begin{equation}\label{mors1}
   \varepsilon_n=\frac{n (M+2)}{M-n+1},
  \qquad \mu(n)= \frac{(M+2)^n
  \Gamma(n+1)\Gamma(M-n+1)}{\Gamma(M+1)}.
\end{equation}
   For the normalization factor in this case one obtains
\begin{eqnarray}\label{mors2}
   \widetilde{\N}(|z|^2)=\left[1+\frac{|z|^2}{2+M}\right]^M,
\end{eqnarray}
   where again the equation (\ref {normD}) has been used.
   Noticing that the series led to $\widetilde{\N}(|z|^2)$ in (\ref {mors2})
   is finite, it is readily found that it is absolutely convergent.
   i.e., $ x=\sqrt{|z|^2}\geq 0, z \in \mathbb{C}$.
   For evaluating the overlap between two of these states, when $\alpha=
   \alpha'$, the formula (\ref{overlap}) is not useful and one must calculate especially
   the overlap between the Morse states for themselves, because of the upper bound of the
   summation
\begin{eqnarray}
     \langle \widetilde{ z, \alpha} |  \widetilde{z', \alpha }\rangle
  &=& \widetilde{\N}(|z|^2)^{-1/2} \widetilde{\N}(|z'|^2)^{-1/2}
     \sum_{n=0}^{M+1} \frac{(z^* z')^n} {\mu(n)}\nonumber\\
  &=&\left[\left(1+\frac{|z|^2}{2+M}\right)
     \left(1+\frac{|z'|^2}{2+M}\right)\right]^{-M/2}
     \left[\frac{2+M+zz'}{2+M} \right]^M.
\end{eqnarray}
   We need only to verify the resolution of the identity. As before, we have to find a
   function $\tilde{\sigma}(x)$ such that
\begin{equation}\label{morse3}
    \int_0^\infty x^n \tilde{\sigma}(x) dx=\frac{(M+2)^n
    \Gamma(n+1)\Gamma(M-n+1)}{\Gamma(M+1)}.
\end{equation}
   Using the definition of Meijer's $G$-function and the inverse Mellin
   theorem, it follows that \cite{Mathai1973}
 \begin{eqnarray}\label{Meiger}
    \int_0^\infty dx x^{s-1} G_{p, q}^{m, n}
     \left( \alpha x \Big|
   \begin{array}{cccccc}
      a_1, & \cdots, & a_n, & a_{n+1}, & \cdots, & a_p  \\
      b_1, & \cdots, & b_m, & b_{m+1}, &\cdots, & b_q
   \end{array}
     \right)  \nonumber\\   =   \frac{1}{\alpha ^s} \frac  {\Pi_{j=1}^{m}\Gamma
     (b_j +s) \Pi_{j=1}^{n}\Gamma (1-a_j-s)} {\Pi_{j=n+1}^{p}\Gamma (a_j+s)
     \Pi_{j=m+1}^{q}\Gamma (1-b_j-s)}.
 \end{eqnarray}
   Comparing (\ref {morse3}) and (\ref {Meiger}),
   one can find the function $\tilde{\sigma}(x)$ needed in (\ref {morse3}), in terms of the
   Meijer's $G$-function by the expression
\begin{eqnarray}\label{Meiger1}
 \tilde{\sigma}(x) = (M+2) \Gamma(M+1) G_{0, 0}^{1, 1}
     \left(  x (M+2)^{-1}\Big|
   \begin{array}{rr}
      -(M+1),& .  \\
        0   ,& .
   \end{array}
      \right).
      \end{eqnarray}
 \vspace{4 mm}\\
    {\bf Example} $5$  {\bf Hydrogen-like spectrum:}
 \vspace{2 mm}\\
    As the final physical example, we choose the Hydrogen-like
    spectrum whose the corresponding CS, has been a long-standing
    subject and discussed frequently in the literature.
    For instance in Refs. \cite{Gazeau1999, Klauder2001} the one-dimensional
    model of such a system with the Hamiltonian $\hat H = -\omega/(\hat n +1)^2$
    and the eigen-values $E_n = -\omega/(n +1)^2$ has been considered ($\omega=m
    e^4/2$, and $n=0, 1, 2, \cdots$).
    But to be consistent with the GKCSs, as it has been done in
    \cite{Klauder2001}, the energy-levels should be shifted by a
    constant amount, such that after taking $\omega=1$ one has the
    eigen-values $e_n$ and therefore the functions $\rho(n)$ as follows
\begin{equation}\label{hyd}
    e_n = 1- \frac{1}{(n+1)^2},
    \qquad \rho(n)=\frac{(n+2)}{2(n+1)}
\end{equation}
    with unit disk centered at the origin as the region of convergence, i.e. $R = 1$.
    Therefore the related dual family of GKCSs for bound state
    portion of the Hydrogen-like atom can be constructed.
    For this purpose, we take into account (\ref{hyd}) in (\ref{relate-dual})
    and (\ref{Hamilt2}), so
    the corresponding quantities for the DGKCSs
    can be easily obtained as
\begin{equation}\label{hyd1}
  \varepsilon_n=\frac{n (n+1)^2}{n+2},
  \qquad \mu(n)= \frac{2 n!(n+1)!}{n+2}.
 \end{equation}
   In this case $\tilde R =\infty$ as the radius of convergence.
   For the normalization factor, using the Eq. (\ref {normD}) one obtains
\begin{eqnarray}\label{hyd2}
   \widetilde{\N}(|z|^2)=
   \frac{1}{2\sqrt{|z|^2}}\left[2I_1(2\sqrt{|z|^2}) + \sqrt{|z|^2}
   I_2(2\sqrt{|z|^2})\right],
\end{eqnarray}
   where $I_\nu(z)$ is the modified Bessel function of the first
   kind.
   Similar to the preceding examples, we only verify the resolution of the identity.
   In the present case we have to find a
   function $\tilde{\sigma}(x)$ such that
\begin{equation}\label{hyd3}
    \int_0^\infty x^n \tilde{\sigma}(x) dx = 2\frac{
    n!(n+1)!}{n+2}.
\end{equation}
   The integral in (\ref {Meiger}) is again helpful, if we rewrite the RHS of the
   (\ref {hyd3}) as $2n![(n+1)!]^2/(n+2)!$. The suitable measure
   is then found to be
 \begin{eqnarray}\label{hyd4}
    \tilde{\sigma}(x) =  G_{0, 0}^{3, 1}
     \left( x \Big|
   \begin{array}{rr}
      0, 1, 1,& .  \\
        2   ,& .
   \end{array}
      \right).
      \end{eqnarray}
   The overlap between these states for the special case  $\alpha= \alpha'$ is
   obtained from (\ref{overlap}) in the closed form
\begin{equation}\label{overlap-hyd}
     \widetilde{ \langle z, \alpha }| \widetilde{z',
     \alpha}\rangle =
      \widetilde{\N}(|z|^2)^{-1/2} \widetilde{\N}(|z'|^2)^{-1/2}
      \frac{1}{2\sqrt{z^\ast z'}}\left(2I_1(2\sqrt{z^\ast z'})+ \sqrt{z^\ast z'}
      I_2(2\sqrt{z^\ast z'})\right),
\end{equation}
   where $\widetilde{\N}(|z|^2)$ and $\widetilde{\N}(|z'|^2)$
   are determined by Eq. (\ref{hyd2}).


\section {\bf Introducing the generalized GKCS and the associated dual family}
    In the light of the above explanations we are now in a position to propose
    the generalized GKCSs, by which we may recover the
    GKCSs given by Eq. (\ref{GKED}) (also the associated dual family, DGKCSs)
    and the nonlinear CSs given by Eq. (\ref{nonl-cs})
    as some special cases. In the following scheme,
    the physical meaning of the $\alpha$ parameter
    which enters in the GKCS and DGKCS will be more clear, the
    case we have already mentioned (in the explanations after Eq. (\ref {transfer}))
    as $\alpha=\omega t$.
\subsection {Time evolved CSs as the generalized GKCSs }
    Consider the Hamiltonian $\hat{H}$ whose eigenvectors and eigenvalues are
    $|\phi_n\rangle$ and $e_n$, respectively such that
\begin{equation}\label{hamlt}
   \hat{H}=\omega \sum_{n=0}^\infty e_n |\phi_n\rangle \langle
   \phi_n|, \qquad \text{where}\quad \hat{H}|\phi_n\rangle=\omega e_n
   |\phi_n\rangle,
\end{equation}
   where $\omega$ is a constant with the dimension of energy (taking
   $\hbar=1$).
   Let $\HH$ be a separable, infinite dimensional and complex Hilbert
   space spanned by orthonormal set $\{|\phi_n
   \rangle\}_{n=0}^\infty$.
   Also suppose $0 = e_0 < e_1 < e_2 < \cdots < e_{n} < e_{n+1} < \cdots
   $, be such that the sum $\sum_{n=0}^\infty \frac {x^n}{[e_n]!}$
   converges in some interval $0 < x \leq L$.
   For $z\in \mathbb{C}$, such that $|z|^2 < L \leq \infty$, we define the generalized CSs
   as follows
\begin{equation}\label{CS}
  |z\rangle \doteq \N(|z|^2)^{-1/2}\sum_{n=0}^\infty
  \frac{z^n}{\sqrt{[e_n]!}}|\phi _n\rangle,
\end{equation}
   where $\N(|z|^2)$ being a normalization factor. As it is clear, these states
   known as nonlinear CSs, with the nonlinearity function $f(n)=\sqrt{\frac{e_n}{n}}$.
   Setting $z=r e^{i\theta}$ with $r=J^\frac{1}{2}$, it is reasonable to
   write $|z\rangle \equiv |J, \theta \rangle$. Now if $d\nu$ be a
   measure which solves the moment problem
\begin{equation}\label{momp}
   \int_0^L J^n d\nu(J) = [e_n]!, \qquad \int_0^L  d\nu(J) = 1,
\end{equation}
   then these CSs satisfy the resolution of the identity
\begin{equation}\label{riden}
  \int_0^L \left[\int_0^{2\pi} |J, \theta \rangle \langle J, \theta |
  \N(J)\frac{d\theta}{2\pi} \right] d\nu(J) = I_{\HH}.
\end{equation}
   The CSs in (\ref{CS}) evolve with time in the manner
\begin{equation}\label{evolv1}
  |z, t \rangle =  e^{-i\hat{H}t}|z \rangle =
  \N(|z|^2)^{-1/2}\sum_{n=0}^\infty \frac{z^n e^{-i\omega e_n
  t}}{\sqrt{[e_n]!}}|\phi _n\rangle,
\end{equation}
    or equivalently in terms of the new variables $J$ and
    $\theta$,
\begin{equation}\label{evolv2}
    |J, \theta, t \rangle = e^{-i\hat{H}t}|J, \theta
    \rangle =  \N(J)^{-1/2}\sum_{n=0}^\infty \frac{J^{n/2} e^{i n  \theta}e^{-i \omega e_n
    t}}{\sqrt{[e_n]!}}|\phi _n\rangle.
\end{equation}
    This larger set of GKCSs, we will call them {\it "generalized GKCSs"},
    defined for all $t$, satisfies the resolution of the identity,
\begin{eqnarray}\label{riden2}
    \int_\mathbb{R} &&\!\!\!\!\!\!\!\!\!\left[\int _0^L \left\{\int_0^{2\pi} |J,
       \theta, t \rangle \langle J, \theta, t |
       \N(J)\frac{d\theta}{2\pi}\right\} d\nu(J)\right] d
       \mu_\mathcal{B}\nonumber\\
   & =& \int_0^L\left[ \int_0^{2\pi} |J, \theta, t \rangle \langle J, \theta, t |
       \N(J)\frac{d\theta}{2\pi}\right] d\nu(J)\nonumber\\
   &=& \int_\mathbb{R} \left[ \int_0^L |J,
       \theta, t \rangle \langle J, \theta, t |
       \N(J) d\nu(J) \right] d \mu_\mathcal{B}\nonumber\\
   &=& \sum_{n=0}^\infty |\phi _n\rangle\langle \phi_n|=
       I_{\HH},
\end{eqnarray}
    where $d \mu_\mathcal{B}$ which is really a functional (not a measure)
    is referred to as the {\it Bohr measure},
\begin{equation}
     \langle\mu_\mathcal{B};f\rangle \doteq
     \lim_{T\rightarrow\infty}\frac{1}{2T} \int_{-T}^ T f(x)dx =
     \int_{\mathbb{R}} f(x) d\mu_\mathcal{B}(x),
\end{equation}
    and $f$ is a suitably chosen function over $\mathbb{R}$. In
    particular, if $f(x)=1$ for all $x$, then
    $\langle\mu_\mathcal{B};f\rangle=1$, so that $\mu_\mathcal{B}$
    resembles a probability measure. Therefore writing the Bohr measure as
    an integral has only a symbolic meaning.

    Setting $t=0$ in the "generalized GKCSs" of Eq. (\ref{evolv1}),
    we shall recover the nonlinear CSs and for $\theta =0$,
    the generalized CSs of Eq. (\ref{evolv2}) reduce to the GKCSs $|J, \alpha
    \rangle$ in (\ref{GKCS})
    with $\alpha\equiv\omega t$, which the latter states satisfy the
    resolution of the identity,
\begin{equation}\label{riden3}
  \int_\mathbb{R} \left[ \int_0^L |J,
        \alpha \rangle \langle J, \alpha |
       \N(J) d\nu(J)\right] d \mu_\mathcal{B}(t) = I_{\HH}.
\end{equation}
  The generalized GKCSs $|J, \theta, t \rangle$ in (\ref{evolv2}),
  satisfy the temporal stability condition and the action identity,
  as well as the continuity in the labels and resolution of the
  identity,
\begin{equation}\label{}
   e^{-i\hat{H}t'}|J, \theta, t \rangle = |J, \theta, t+t' \rangle,
   \qquad \langle J, \theta, t |\hat{H}|J, \theta, t \rangle=\omega
   J,
\end{equation}
   and so do the states $|z, t\rangle$ in (\ref{evolv1}).


\subsection {\bf The dual family of the "generalized GKCS"}
   Let us now write $e_n=nf^2(n)$, so using our previous results in
   the present paper,
   there is a {\it dual set} of numbers
   $\widetilde{e}_n \equiv \varepsilon_n=\frac{n}{f^2(n)}$,
   associated with the {\it dual Hamiltonian} $\widetilde{\hat{H}}$.
   Correspondingly this Hamiltonian has eigenvectors
   $|\phi_n\rangle$ and eigenvalues $\varepsilon_n$, such that
\begin{equation}\label{hamlt}
  \widetilde{\hat{H}}=\omega \sum_{n=0}^\infty \varepsilon_n |\phi_n\rangle \langle
  \phi_n|, \qquad \text{where}\quad \widetilde{\hat{H}}|\phi_n\rangle=\omega \varepsilon_n
  |\phi_n\rangle.
\end{equation}
    Also assuming that  $0 = \varepsilon_0 < \varepsilon_1 < \varepsilon_2 <
    \cdots < \varepsilon_{n} < \varepsilon_{n+1} <
    \cdots$, be such that the sum $\sum_{n=0}^\infty \frac {x^n}{[\varepsilon_n]!}$
    converges in some interval $0 < x \leq \widetilde{L}$. We can
    now define the {\it dual family} of generalized CSs have been introduced in (\ref{CS}) by
\begin{equation}\label{CS-dual}
   |\widetilde{z}\rangle \doteq \widetilde{\N}(|z|^2)^{-1/2}\sum_{n=0}^\infty
   \frac{z^n}{\sqrt{[\varepsilon_n]!}}|\phi _n\rangle,
\end{equation}
   which are the well known(dual) nonlinear CSs of Ref. \cite{Roy-Roy2000}.
   The time evolution of these states reads as,
\begin{equation}\label{evolv3}
  |\widetilde{z, t} \rangle =  e^{-i\hat{H}t}|\widetilde{z} \rangle =
  \widetilde{\N}(|z|^2)^{-1/2}\sum_{n=0}^\infty \frac{z^n e^{-i\omega \varepsilon_n
  t}}{\sqrt{[\varepsilon_n]!}}|\phi _n\rangle.
\end{equation}
   Again, setting $z=r e^{i\theta}$ with $r=J^\frac{1}{2}$, we can
   write $|\widetilde{z}\rangle \equiv |\widetilde{J, \theta
   }\rangle$. So, equivalently the states in (\ref {evolv3}) can be
   rewritten in terms of the new variables $J$ and $\theta$ as
\begin{equation}\label{evolv4}
    |\widetilde{J, \theta, t }\rangle = e^{-i\hat{H}t}|\widetilde{J,
    \theta}  \rangle =  \widetilde{\N}(J)^{-1/2}\sum_{n=0}^\infty \frac{J^{n/2}
    e^{i n  \theta}e^{-i \omega \varepsilon_n
    t}}{\sqrt{[\varepsilon_n]!}}|\phi _n\rangle.
\end{equation}
  We call this large set of states as the {\it "dual of the generalized GKCS"}.
  Setting $\theta =0$ in (\ref{evolv4}) will reduce it to the dual
  of the GKCSs we introduced in (\ref{DGKED}) with $\alpha=\omega t$.
  Provided the moment problem
\begin{equation}\label{momp}
  \int_0^{\widetilde{L}}
   J^n d\widetilde {\nu}(J) = [\varepsilon_n]!, \qquad \int_0^{\widetilde
  {L}} d\widetilde{\nu}(J) = 1,
\end{equation}
   has a solution, we also have expressions for the resolution of
   the identity of the type (\ref{riden}), (\ref{riden2}) and
(\ref{riden3}).
   The GK criteria may immediately be verified for the dual of the generalized
   GKCS in Eq. (\ref{evolv4}),
   as it was down for the "generalized GKCSs" in Eq. (\ref{evolv2}).


\subsection {\bf Generalized creation and annihilation operators}
  We define two sets of the generalized annihilation operators
\begin{equation}\label{}
  A|\phi_n\rangle = \sqrt{e_n} |\phi_{n-1}\rangle,
  \qquad \widetilde{A}|\phi_n\rangle = \sqrt{\varepsilon_n}
  |\phi_{n-1}\rangle,
\end{equation}
  and the corresponding generalized creation operators
\begin{equation}\label{}
  A^\dag|\phi_n\rangle = \sqrt{e_{n+1}} |\phi_{n+1}\rangle,
  \qquad \widetilde{A}^\dag|\phi_n\rangle = \sqrt{\varepsilon_{n+1}}
  |\phi_{n+1}\rangle,
\end{equation}
   where  $\varepsilon_n\equiv\widetilde{e}_n$.
   Therefore, the Hamiltonian of the system and its associated dual are
\begin{equation}\label{}
  \hat{H}=\omega A^\dag A, \qquad \widetilde{\hat{H}}=\omega  \widetilde{A}^\dag
  \widetilde{A}.
\end{equation}
   Note that we have dropped the GK indices from all the operators
   in this last section because the discussion is exclusively
   related to the GK states.
   Then, for the states in (\ref{CS}) and (\ref{CS-dual}) we have
   clearly
\begin{equation}\label{}
   A|z\rangle =z|z\rangle, \qquad  \widetilde A|\widetilde z\rangle =z|\widetilde
   z\rangle.
\end{equation}
   In the Heisenberg picture the generalized annihilation operators $A$ and $\widetilde A$
   evolve in time as
\begin{equation}\label{}
  A(t)=e^{i \hat{H} t}A e^{- i \hat{H} t}, \qquad
  \widetilde A(t)=e^{i \widetilde{\hat{H}} t} \widetilde A e^{- i \widetilde{\hat{H}} t}
\end{equation}
   and similarly for the generalized creation operators. At any time
   $t$ one may obtain,
\begin{equation}\label{}
  A(t)|z, -t\rangle =z|z, -t\rangle, \qquad
  \widetilde A(t)|\widetilde {z, -t}\rangle = z |\widetilde{z, -t}\rangle.
\end{equation}

\subsection {\bf Interpolating between generalized GKCSs and their dual}
  Following the approach proposed in \cite{Ali2004} we can define
  the operator $\hat{T}$ on $\HH$ as
\begin{equation}\label{}
   \hat{T} \doteq \sum_{n=0}^\infty \sqrt {
   \frac{[e_n]!}{[\varepsilon_n]!}}|\phi _n\rangle\langle \phi_n|,
\end{equation}
with the action
\begin{equation}\label{}
  \hat{T} |\phi _n\rangle = \sqrt { \frac{[e_n]!}{[\varepsilon_n]!}}|\phi _n\rangle,
   \qquad n=0, 1, 2,  \cdots.
\end{equation}
 Then, writing $\eta _z=\N(|z|^2)^{1/2}|z\rangle$ and  similarly
 defining $\widetilde \eta _z$, $\eta _{J, \theta, t}$ and
  $\widetilde \eta _{J, \theta, t}$ as the unnormalized CSs, we
  have
\begin{equation}\label{}
  \hat{T} \eta_z= \widetilde \eta_z.
\end{equation}
  Since $\hat{T}$, $\hat{H}$ and  $\widetilde{\hat{H}}$ commute, we
  have the interpolation rule at any fixed time $t$; i.e., fixed $\alpha$:
\begin{equation}\label{}
  e^{-i (\widetilde{\hat{H}}-\hat{H})t}\hat{T} \eta_{J, \theta, t}= \widetilde
  \eta_{J, \theta, t}.
\end{equation}

   \section {\bf Concluding remarks}
      Finally we present a summary of our results.
    After imposing a second modification on GK states,
    we showed that both the GKCSs and DGKCSs
    are essentially of the type of the so called nonlinear ($f$-deformed) CSs.
    In each of the two cases the relevant nonlinearity function is an
    {\it operator valued function} which depends on the intensity of light ($\hat{n}$),
    but is labeled by a constant real parameter ($\alpha$).
    The introduced nonlinearity function which
    contains an intensity dependent phase factor, has not been
    appeared in literature up to now.
    This new feature originates from the temporal stability
    requirement imposed on GKCSs and DGKCSs.
    Meanwhile, using the two
    nonlinearity functions we constructed the raising and lowering
    operators, by which one can creates and annihilates
    the {\it deformed photons}.
    After all, we proposed a general evolution operator $\hat
    S(\alpha)$,
     whose action on any generalized CS with known spectrum
     transforms it to a temporally stable CS.
     This can {\it physically} makes the generalized CS to be more
     useful in practical experiments.

    Adding the results presented in section (\ref {sec-Examples})
    show that, at least in most of the considered physical systems,
    i.e., when the CS deals with the whole Fock space
    $\{|n\rangle\}_{n=0}^\infty$, while the GKCSs is defined on
    whole complex plane (unit disk) the DGKCSs
    is restricted to  unit disk (whole plane), and vice versa.
    This situation does not hold for Morse potential in which there
    is a cut off in the summation (finite dimensional
    Hilbert space: $\{|n\rangle\}_{n=0}^{M+1}$).
    So based on the
    results in \cite{Naderi2004a} in which the authors relate the radius of
    convergence to the physical quantities, it can be conclude that
    GKCSs and DGKCSs can be produced under different physical conditions.

     We emphasize on the fact that the
     Hamiltonian involved in the operator $\hat{S}(\alpha)$ must be the one
     that expresses the dynamics of the system.
     Using this proposition we introduced the dual family of GKCS.
     Also it may be understood that the GKCSs can be rewritten in terms of
     the associated $f_{GK}$ function, explicitly as
     \[|z, \alpha\rangle_{f} = \N(|z|^2)^{-1/2}\sum_{n=0}^\infty \frac{z^n}
     {\sqrt{n!}[f_{GK}(\alpha, \hat{n})]!}|n\rangle, \]
     and similarly for the DGKCSs in terms of the same nonlinearity function
     $f_{GK}(\alpha, \hat n)$:
     \[|\widetilde{z, \alpha} \rangle_{f} = \widetilde{\N}(|z|^2)^{-1/2}\sum_{n=0}^\infty
     \frac{z^n  [f_{GK}(\alpha, \hat{n})]!}{\sqrt{n!}}|n\rangle. \]
     Hence, we have established that the latter states are indeed a special class of
     nonlinear CSs which are temporally stable. This property is
     preserved, using a particular set of nonlinearity functions
     as introduced in (\ref {nlGKED}) and (\ref {nlDGKED}).
     Also, the map
     $$ |z, \alpha\rangle \mapsto |\widetilde {z, \alpha}\rangle$$
     may be obtained by following map
     $$ f_{GK}(\alpha, \hat{n}) \mapsto  \frac {1} {f_{GK}(\alpha, \hat{n})}.$$
     Note that this map converts even the normalization factor,
     correctly. Keeping in mind the above explanation, a certain class of nonlinear CS
     and its dual, determined on a specified point of the phase space ($z \in \C$),
     including the fact that the standard CS is self-dual, may be similitude
     to the image of a state by the map defined above.
     While the images of standard canonical CS are the same
     in any arbitrary point of the phase space,
     i.e., the usual CS does not destroy the flatness of the
     mirror (or the linearity of the medium),
     this is not so for the nonlinear CSs.
     It may be recognized that the operator $f$ in the CSs
     affect the flatness of the mirror (or the linearity of the medium),
     and makes it to be curved (or nonlinear).
     As much as the state is far from  the harmonic
     oscillator CSs, the effect of $f$-function is more strong and the dual state(image)
     is far from the state, itself. So the role of the
     nonlinearity in the nonlinear CSs
     may be related to the medium, instead of the
     source of light. In other words, corresponding to any nonlinear
     CSs there exist an equivalent situation: "a linear source(ordinary photon)
     with a {\it nonlinear optical medium}", such as Kerr medium.
     This particular interpretation
     of the nonlinear CSs has already been realized in
     \cite{Mizrahi2004, Naderi2004b}.

     Finally, we confined ourselves to the general introduction of the new
     type of Gazeau-Klauder CSs, in this paper.
     Although it would be interesting for further applications in quantum optics,
     to investigate in details,
     minimization of Robertson-Schr\"{o}dinger uncertainty
     equation, the intelligent states and the statistical properties, quadrature
     squeezing,
     {\it etc} of DGKCSs and the special superpositions of them have been introduced
     associated with various exactly solvable quantum mechanical
     systems and compared with those of the old one (GKCSs type).
     Also it seems that the vector CSs of DGKCSs may be constructed,
     as it was done for GK states \cite{ali-bag}.
     These matters are under consideration for future works.
\begin{acknowledgments}
   The authors would like to express their utmost thanks to Prof. S. Twareque Ali
   from the Department of Mathematics and
   Statistics of Concordia University, for his intuitive and useful
   comments and suggestions,
   in addition to his rigorous look at the manuscript, in all
   its stages during preparation.
   Thanks to Prof. M. Soltanolkotabi and Dr. M. H. Naderi
   for reading carefully the manuscript and useful discussions.
   Also thanks to the referee for reminding us some useful comments.
   One of the authors (M.K.T.) acknowledges supports from the University of Yazd of Iran.
\end{acknowledgments}


\end{document}